# Tunneling inhibition for subwavelength light

Changming Huang,[1] Xianling Shi,[1] Fangwei Ye,[1,*] Yaroslav V. Kartashov,[2,3] Xianfeng Chen[1] and Lluis Torner[2]

[1]*State Key Laboratory of Advanced Optical Communication Systems and Networks, Physics Department, Shanghai Jiao Tong University, Shanghai 200240, China*
[2]*ICFO-Institut de Ciencies Fotoniques, and Universitat Politecnica de Catalunya, 08860 Castelldefels (Barcelona), Spain*
[3]*Institute of Spectroscopy, Russian Academy of Sciences, Troitsk, Moscow Region 142190, Russia*
*\*Corresponding author: fangweiye@sjtu.edu.cn*

We show that light tunneling inhibition may take place in suitable dynamically modulated waveguide arrays for light spots whose features are remarkably smaller than the wavelength of light. We found that tunneling between neighboring waveguides can be suppressed for specific frequencies of the out-of-phase refractive index modulation, affording undistorted propagation of the input subwavelength light spots over hundreds of Rayleigh lengths. Tunneling inhibition turns out to be effective only when the waveguide separation in the array is above a critical threshold. Inclusion of a weak focusing nonlinearity is shown to improve localization. We analyze the phenomenon in purely dielectric structures and also in arrays containing periodically spaced metallic layers.
http://dx.doi.org/10.1364/OL.99.099999

One of the most exciting directions in modern optics is the development of new strategies for controlling the propagation path and diffraction rate of the eigenmodes that remain invariant on propagation or evolve in a desired fashion. Very frequently such strategies rely on the use of artificial composite materials having spatially inhomogeneous refractive index [1]. Thus, an unprecedented freedom in the engineering of spatial dispersion afforded by periodic structures such as photonic crystals and coupled waveguide arrays, allows observation of many intriguing phenomena that do not occur in natural uniform materials [2]. Among such phenomena is the possibility to change the strength and sign of diffraction, as was demonstrated in straight waveguide arrays or in those which bend periodically along the propagation path [3,4]. In general, a broad spectrum of periodic variations of the refractive index may induce resonant cancellation of diffractive broadening [5]. This physical effect is reminiscent of the arrest of wavepacket tunneling by external driving fields, studied in electronic systems [6,7]. In optics, two manifestations of this effect have been observed experimentally. Namely, dynamic localization in periodically curved arrays [8-14] and inhibition of tunneling in straight arrays with an out-of-phase modulation of the longitudinal refractive index in the neighboring channels [15-21].

However, all these schemes for allowing the control of the rate of diffractive broadening have only been realized in the paraxial regime, when all characteristic scales, such as the beam width and array period, substantially exceed the wavelength of light. In this regime, light propagation can be modeled by the scalar paraxial Schrödinger equation that also describes the evolution of excitations in a number of other physical systems besides classical optics [22].

At the same time, the development of approaches for diffraction control at the subwavelength scale may open up important opportunities for the miniaturization of photonic devices for confining and manipulating light. Rapidly developing nanofabrication techniques already allow the creation of optical structures with characteristic scales much smaller than the wavelength of light. The large diffraction angle inherent to the subwavelength light spots excludes the use of the paraxial approximation for the description of their propagation in such structures. The coupling of transverse and longitudinal field components can no longer be ignored for such beams and one must take into account the vectorial nature of light. Therefore, a salient question arises: can dynamically varying guiding structures be used for the control and inhibition of strong diffraction of subwavelength beams?

In this letter, using the solution of the full set of Maxwell's equations, we show that resonant inhibition of tunneling can be achieved even for subwavelength light spots in purely dielectric (or metallic-dielectric) periodic structures where both the width of the waveguides and their separation are smaller than the wavelength of the light. We found that a necessary ingredient of tunneling inhibition is an out-of-phase longitudinal refractive index modulation in neighboring guiding channels. The rate of diffraction broadening of light beams in such a structure is controlled by the detuning of modulation frequency from one of the resonant values at which the subwavelength spot remains confined in the excited guide. We show how downscaling of the entire structure affects the effectiveness of the tunneling inhibition.

We start our analysis by considering the propagation of a TM-polarized light beam along the $z$-axis of a dielectric material with transversally and longitudinally modulated refractive index. The evolution of the components of the electric and magnetic fields $(E_x; H_y; E_z)$ that remain nonzero for selected polarization state is governed by the reduced system of Maxwell's equations,

$$\begin{aligned} i\frac{\partial E_x}{\partial z} &= -\frac{1}{\varepsilon_0 \omega}\frac{\partial}{\partial x}\left(\frac{1}{\varepsilon_r}\frac{\partial H_y}{\partial x}\right) - \mu_0 \omega H_y, \\ i\frac{\partial H_y}{\partial z} &= -\varepsilon_0 \varepsilon_r \omega E_x, \end{aligned} \quad (1)$$

where the longitudinal component of the electric field $E_z = (i/\varepsilon_0 \varepsilon_r \omega)\partial H_y/\partial x$ was excluded from (1) for convenience, $\varepsilon_0$ and $\mu_0$ are the vacuum permittivity and permeability, $\omega$ is the frequency of light, $\varepsilon_r(x,z)$ is the relative permittivity of the underlying structure, whose shape in the case of linear dielectric material is described by the function

$$\varepsilon_r(x,z) = \varepsilon_{\text{bg}} + p\sum_{m=-M}^{+M}[1+(-1)^m\delta\sin(\omega_z z)]\exp[-(x-md)^2/a^2], \quad (2)$$

where $\varepsilon_{\text{bg}}$ is the relative background permittivity, $\delta$ is the longitudinal modulation depth, the parameters $a, p$ characterize the width and depth of Gaussian waveguides, $d$ is the separation between neighboring waveguides, and $\omega_z$ is the frequency of modulation of the permittivity in the longitudinal direction (below we will normalize it by the frequency $\omega_b$ of power switching between two unmodulated guides).

We fix the wavelength of the light beam $\lambda = 632.8$ nm, select $\varepsilon_{\text{bg}} = 2.25$, $p = 1.7$, and consider subwavelength waveguides with the width of $a = 100$ nm $\ll \lambda$ and separation $d = 700$ nm or even smaller. We used at the input the eigen-mode of the isolated waveguide that can be obtained as a stationary solution $[E_x(x,z), H_y(x,z)] = [E_x(x)e^{i\beta z}, H_y(x)e^{i\beta z}]$ of Eqs. (1) with $\delta = 0$. The waveguide width and depth were adjusted such that it supports only the guided mode whose shape is shown in Fig. 1(a). One can see that the longitudinal field component is comparable in amplitude with the transverse field component. The exponential tails of such a mode extend beyond a narrow subwavelength waveguide, but its integral width remains well below $d = 700$ nm. If such a mode is launched into the system of two unmodulated waveguides one observes periodic power exchange between channels with the frequency $\omega_b \approx 5.6 \times 10^4$ m$^{-1}$. Further reduction of the waveguide width down to $50$ nm does not lead to a pronounced decrease of the mode width (due to the diffraction limit) even if the depth $p$ of the waveguide is increased so as to keep the same beating frequency $\omega_b$.

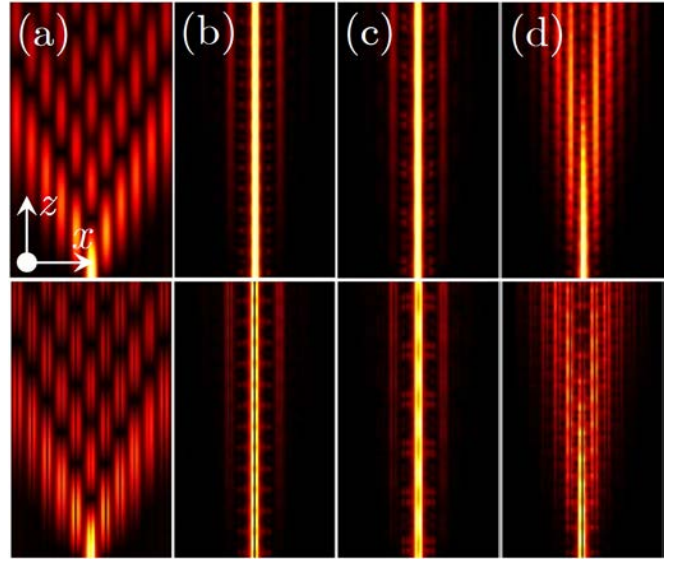

Fig. 2. The evolution of the modulus of the transverse $E_x$ (top row) and longitudinal $E_z$ (bottom row) electric field components for modulation frequencies $\omega_z/\omega_b = 0$ (a), $\omega_z/\omega_b = 7.4$ (b), $\omega_z/\omega_b = 3.19$ (c), and $\omega_z/\omega_b = 6$ (d). In all cases the propagation distance is $200\ \mu$m, while the transverse width of the depicted window is $7.7\ \mu$m.

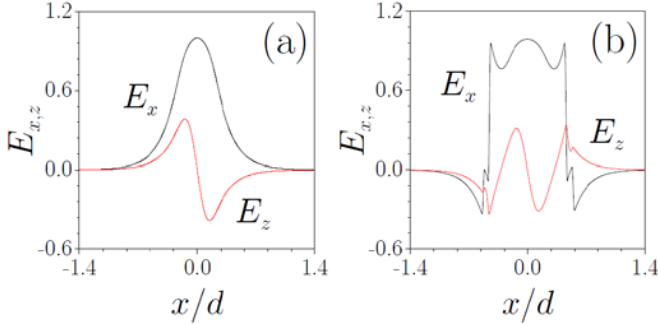

Fig. 1. Transverse and longitudinal electric field distributions in linear modes supported (a) by the purely dielectric waveguide when $\varepsilon_r(x) = \varepsilon_{\text{bg}} + p\exp(-x^2/w^2)$ for $x/d \in (-\infty, +\infty)$, and (b) by the dielectric waveguide sandwiched between semi-infinite metal layers when $\varepsilon_r(x) = \varepsilon_{\text{bg}} + p\exp(-x^2/w^2)$ for $|x/d| \leq 0.464$ and $\varepsilon_r(x) = \varepsilon_m$ for $|x/d| > 0.464$.

Figure 2(a) illustrates the propagation dynamics in an unmodulated waveguide array when only a single channel is excited at $z = 0$. One can see that already at $z = 100\ \mu$m light expands over approximately ten waveguides, i.e., the diffraction angle is considerable despite the transverse refractive index modulation. The characteristic "discrete diffraction" pattern is observed for both the transverse and longitudinal electric field components. This picture changes significantly if the waveguides feature out-of-phase longitudinal modulation with $\delta = 0.2$ at properly selected frequencies $\omega_z$. In this case the tunneling to neighboring waveguides due to the overlap of the tails of their guided modes is almost completely inhibited and the light remains confined in the excited waveguide as shown in Figs. 2(b) and 2(c), which correspond to the primary and secondary resonant frequencies [see Fig. 3(a)]. Notice that our finite-element method takes into account backward reflection, which was negligible in all cases considered. The inhibition of tunneling occurs because out-of-phase refractive index modulation results in a renormalization of the effective coupling constant $\kappa_{\text{eff}} = \kappa J_0(\mu\delta/\omega_z)$ (here $\kappa \sim \omega_b$ is the coupling constant in the unmodulated array) characterizing the rate of power exchange between the waveguides. This can be shown in a "discretized" version of Eq. (1) operating with modal amplitudes in coupled guides (under the assumption of weak coupling, the derivation procedure for the nonparaxial case leads to discrete NLSE [18]). The possibility of inhibiting the strong diffraction of the nonparaxial light spots by longitudinal refractive index modulations is one of the central results of this letter.

The effectiveness of the tunneling inhibition can be characterized by the dependence of the distance-averaged power fraction trapped in the excited channel

$$U_{\text{av}} = (UL)^{-1}\int_0^L dz \int_{-d/2}^{+d/2}(|E_x(x,z)|^2 + |E_z(x,z)|^2)dx,$$
$$U = \int_{-d/2}^{+d/2}(|E_x(x,0)|^2 + |E_z(x,0)|^2)dx, \quad (3)$$

on the frequency of the longitudinal modulation $\omega_z$ shown in Fig. 3(a) for the case of $L = 200\ \mu$m. One observes several resonance spikes, with the primary reson-ance corresponding to the highest modulation frequency. The density of the resonances grows with a decrease of $\omega_z$. The inhibition is most effective in the primary resonance, where for the parameters of our array, $U_{av} \approx 0.94$. The propagation dynamics corresponding to the primary and secondary resonances is illustrated in Figs. 2(b) and 2(c), while Fig. 2(d) shows the slow beam broadening for the off-resonant modulation frequency.

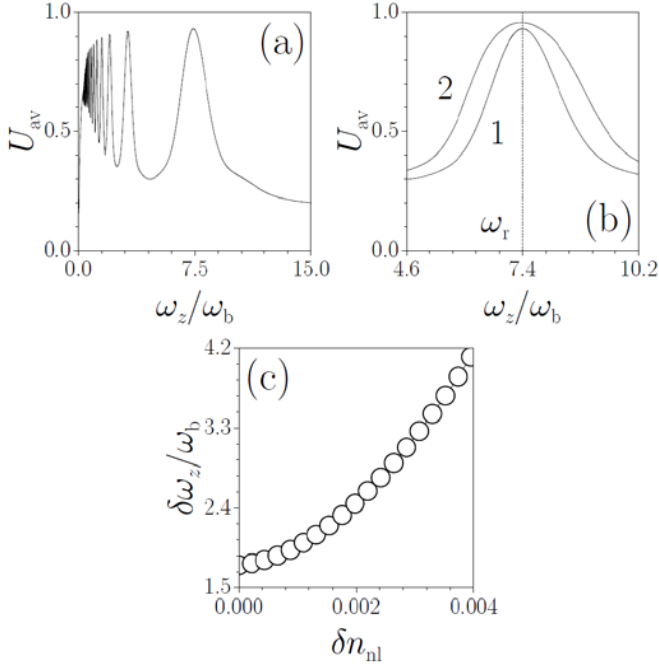

Fig. 3. (a) Distance-averaged power in the excited waveguide versus $\omega_z / \omega_b$ in the linear waveguide array. (b) $U_{av}$ versus $\omega_z / \omega_b$ at $\delta n_{nl} = 0$ (curve 1) and $\delta n_{nl} = 0.0025$ (curve 2). (c) The width of the primary resonance as a function of the nonlinear contribution to the refractive index. In all cases $z = 200$ $\mu$m.

A focusing nonlinearity of the medium further enhances the tunneling inhibition. In order to study the impact of nonlinearity, we assume the presence of a Kerr contribution to the dielectric permittivity $\varepsilon_r$ proportional to the total field intensity $|E_x|^2 + |E_z|^2$. The increase of peak nonlinear contribution $\delta n_{nl}$ to the refractive index results in a simultaneous broadening of all resonances in $U_{av}(\omega_z)$ dependence (this effect is illustrated in Fig. 3(b) for primary resonance), while resonant frequencies are not affected by the nonlinearity. Remarkably, due to the considerable nonlinearity-induced broadening of the resonances, one can achieve a nearly complete inhibition of tunneling for non-resonant modulation frequencies at very low power levels. In Fig. 4 we show how diffraction is replaced by localization upon increase of the nonlinear contribution to the refractive index up to $\delta n_{nl} \approx 0.003$ in the case when the modulation frequency $\omega_z$ is detuned by approximately 10% from the frequency of the primary resonance. This nonlinear contribution is exceptionally small in comparison with the $\delta n_{nl}$ values required for the excitation of subwavelength solitons in unmodulated systems [23-29]. Therefore, longitudinal refractive index modulation dramatically reduces the thresholds for excitation of localized nonlinear modes. We found that an increase of the nonlinear contribution to the refractive index results in a monotonic growth of the resonance width $\delta \omega_z$ defined at the level $0.7 \max(U_{av})$ [Fig. 3(c)]. The value $\delta \omega_z$ becomes comparable with the resonance frequency already at $\delta n_{nl} = 0.004$.

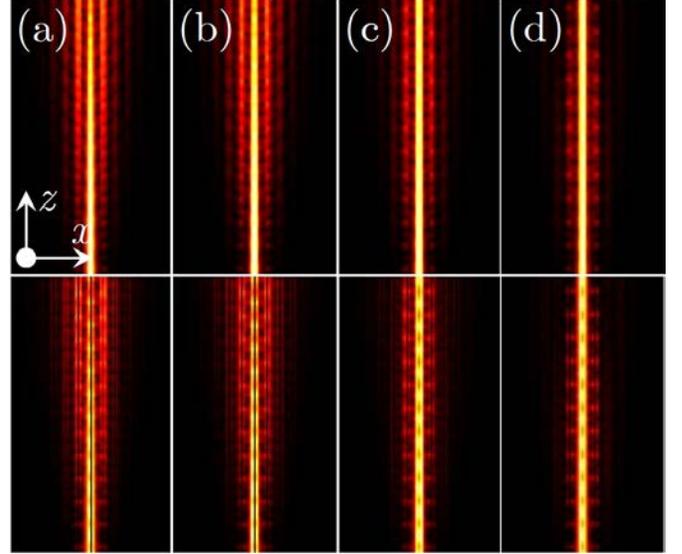

Fig. 4. The evolution of $|E_x|$ (top row) and $|E_z|$ (bottom row) upon propagation in the nonlinear modulated waveguide array for $\delta n_{nl} = 0$ (a), $\delta n_{nl} = 0.00063$ (b), $\delta n_{nl} = 0.00188$ (c), and $\delta n_{nl} = 0.00301$ (d). In all cases $\omega_z / \omega_b = 6.7$, while the transverse and longitudinal scales are the same as in Fig. 2.

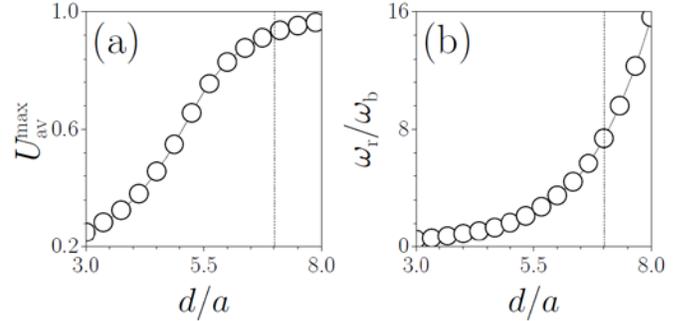

Fig. 5. Maximal distance-averaged power (a) and primary resonance frequency (b) versus separation between waveguides. In all cases the propagation distance is $z = 200$ $\mu$m. Dashed line corresponds to separation used in Figs. 1-4.

As mentioned above, the subwavelength mode that we used for the excitation of the central waveguide is nearly diffraction-limited and further reduction of the waveguide width $a$ at a fixed period $d$ does not qualitatively change the dynamics of the tunneling inhibition if one adjusts the waveguide depths to maintain a fixed coupling strength. Further miniaturization of our system can be achieved at the expense of a reduction in the separation between the waveguides. The dependence on this separation of the maximal distance-averaged power fraction in the excited channel achieved in the primary resonance is shown in Fig. 5(a). In simulations we tuned the waveguide depth $p$ in order to get the same beating frequency $\omega_b$ for different $d$ values. This allows direct comparison of the effectiveness of the tunneling inhibition for different $d$ because identical $\omega_b$ imply an equal diffraction strength in the different arrays. While for $d/a \geq 7$ the tunneling inhibition is almost equally strong for all array periods, its effectiveness rapidly drops already for $d/a \sim 5$ indicating the existence of a minimum scale below which longitudinal modulation cannot compensate for diffractive broadening. The scaled primary resonance frequency grows with increasing $d$ [Fig. 5(b)].

Finally, we found that inhibition of tunneling is possible not only in purely dielectric subwavelength structures, but also in the case when narrow $50$ nm metallic layers are introduced between $100$ nm -wide Gaussian waveguides separated by the distance $d=700$ nm. As before we suppose that refractive index of dielectric Gaussian waveguides is modulated in longitudinal direction. The permittivity of metal (silver) $\varepsilon_{\mathrm{m}} \approx -20 - 0.19i$ was calculated using Drude model. Initially we neglect the losses in the metal and assume that $\mathrm{Im}\,\varepsilon_{\mathrm{m}}=0$. The example of eigenmode of an unmodulated Gaussian waveguide sandwiched between two semi-infinite metal layers that was used as an input in simulations of the propagation is shown in Fig. 1(b). The metal strongly alters the shape of the mode and leads to sharp field variations near the interface with the dielectric. Still, the modal field considerably penetrates into the metal region, leading to the possibility of an evanescent coupling between neighboring waveguides. This coupling is illustrated in Fig. 6(a) for unmodulated dielectric-metal array of waveguides. The inhibition of tunneling in the primary resonance is shown in Fig. 6(b), while the propagation for an off-resonant frequency is shown in Fig. 6(d). Notice that in the presence of metal, the radiation emitted from the central waveguide is more pronounced. Since the metallic stripes are narrow, the Ohmic losses inherent to metallic structures do not affect the tunneling inhibition, Fig.6(c).

In conclusion, we showed that longitudinal modulation of the parameters of subwavelength waveguide arrays can lead to resonant suppression of diffraction, *even for fully nonparaxial light spots*. Such modulation drastically reduces the thresholds for the formation of nonlinear excitations. This inhibition of tunneling can be effective only when the period of the array exceeds a critical value. Our results open up the possibility to control light at subwavelength scales in unprecedented ways. From a practical point of view, such control may lead to significant miniaturization of future photonic devices for signal processing.

The work of C. Huang and F. Ye was supported by the NSFC, Grant No. 11104181.

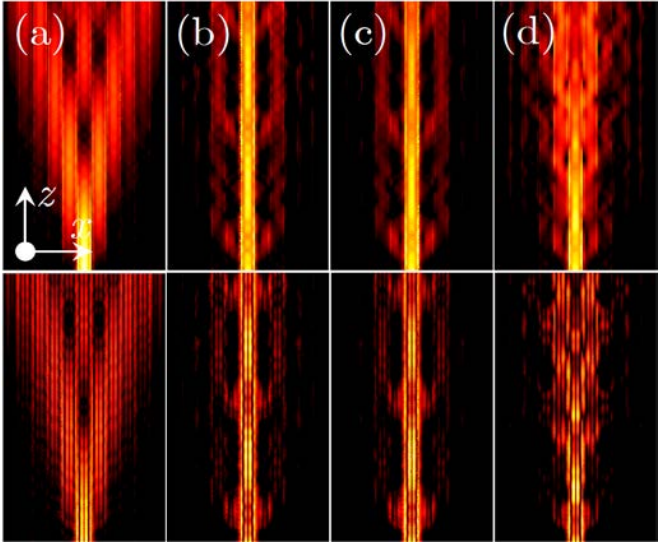

Fig. 6. Inhibition of tunneling in dielectric-metal waveguide arrays. The top row shows $|E_x|$, while the bottom row shows $|E_z|$. (a) Unmodulated array. Modulated array with $\omega_z/\omega_b = 2.09$ without losses in metal (b) and with losses in metal (c). (d) Modulated array with $\omega_z/\omega_b = 3$. In all cases the propagation distance is $50$ $\mu$m, while the transverse width of the depicted window is $7.7$ $\mu$m.